\documentclass[aps,pra,twocolumn]{revtex4}   

\newcommand {\be}{\begin{equation}}
\newcommand {\ee}{\end{equation}}
\newcommand {\bey}{\begin{eqnarray}}
\newcommand {\eey}{\end{eqnarray}}
\usepackage{epsfig}

\begin{document}

\title{Bistability and macroscopic quantum coherence in a BEC of $^7Li$.}

\author{A.~Montina $^*$ and F.T.~Arecchi $^\dagger$  \\
$^*$Dipartimento di Fisica, Universit\`a di Pisa,
Via F. Buonarroti 2, \\ 
56127 Pisa, Italy\\
$^\dagger$ Dipartimento di Fisica, Universit\`a di Firenze \\and\\
INOA (Istituto Nazionale di Ottica Applicata), Largo E.~Fermi 6,50125 Firenze,
Italy}
\date{\today}

\begin{abstract}

We consider a Bose-Einstein condensate of $^7Li$ in a situation where the density 
undergoes a symmetry breaking in real space. This occurs for a suitable 
number of condensed atoms in a double
well-potential, obtained by adding a standing-wave light field to the trap potential.
Evidence of bistability results from the solution of the Gross-Pitaevskii equation.
By second quantization, we show that the classical bistable situation leads, in fact,
to a macroscopic quantum superposition or Schr\"odinger cat (SC) and 
evaluate the tunneling rate between the two SC states. The oscillation between the 
two states is called macroscopic quantum coherence (MQC); we study the effects of 
losses on MQC.
\end{abstract}

\maketitle
\section{Introduction}

The availability of Bose-Einstein condensates of trapped cold 
atoms~\cite{RbRef,Ketterle,LiRef,Li97}
has opened the possibility of a laboratory engineering of quantum 
states with a large number of atoms~\cite{Walsworth} (around a thousand for 
$^7Li$~\cite{Li97,RbCond}).

One of the most challenging endeavours of quantum engineering is the evidence of 
superposition states [so called SC=Schr\"odinger cat (SC)] whose mutual interference 
is called macroscopic quantum coherence (MQC). SC have been observed, e.g., for states of 
a trapped 
ion \cite{Monroe} and of a microwave field in a high-{\cal Q} cavity \cite{Brune}.

In this paper we demonstrate the reliable preparation of SC consisting of a 
Bose-Einstein condensate (BEC) of $^7Li$ 
atoms that have negative scattering length, trapped in a double-well potential. 
Realistic calculations have been offered for macroscopic quantum 
tunneling (MQT) \cite{Leg,Sala}. Indeed, combining the kinetic and potential terms of 
a harmonic 
trap with the inter-particle attraction yields a metastable state for $N<N_c$ ($N_c=$ 
critical population for an attractive BEC). Quantum tunneling from this metastable 
state towards the collapsed state, which would otherwise be reached for $N>N_c$, has 
been shown to be feasible. 

A BEC of atoms with negative scattering length, 
trapped in a double 
well, undergoes a space symmetry breaking beyond a threshold number of atoms $N_i$, whereby
two stable states are formed.
This phenomenon has been dealt with theoretically by a two mode approach~\cite{Ho}.

We have studied the problem by finding numerically the stationary solutions of the 
Gross-Pitaevskii (GP) equation discretized over a space lattice, with reference to the
$^7Li$ case~\cite{artipre}.
Once the stationary solutions have been found, we introduce a quantum two mode model,
with the two modes chosen in such a way as to reproduce the stationary solutions of GP.
The model shows the feasibility of macroscopic quantum coherence (MQC).

Bistability occurs only for an attractive interatomic potential (negative scattering
length). 
Two proposals for reaching SC in a BEC with repulsive atoms have been put 
foward~\cite{Cirac,Gordon}. They both require a Raman coupling between two different 
components; Both the papers consider
copropagating light beams. In Ref.~\cite{Cirac}, MQC is shown to require values of the
scattering length between atoms of different magnetic number substantially
larger than the scattering length between atoms of equal magnetic number. This
requirement is too strong, since
no experimental technique is today accessible to provide such a difference;
furthermore, if such a difference could be achieved, outstanding symmetry-breaking
effects would occur~\cite{Savage}. Reference~\cite{Gordon} introduces a time-dependent 
evolution, so that SC is reached over a time of the order of 1 sec.
However, no clear-cut experimental test is offered to discriminate between SC
and a statistical mixture of two separate states.

\section{Bistability and symmetry breaking}

\begin{figure}
\epsfig{figure=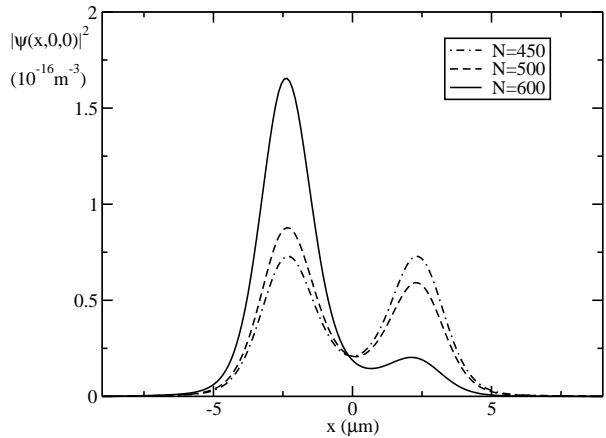,width=7cm,angle=-90}
\caption{Density distribution of the lithium BEC in real space for $N=450$ (dash-dotted line),
$N=500$ (dashed line), and $N=600$ (solid line) with $A_n\equiv A/\hbar=2650 s^{-1}$ and
$\sigma=5 \mu m$.}
\label{fig1}
\end{figure}

We refer to a condensate of $^7Li$ atoms trapped in a double-well 
potential. A suitable model for it is given by 
\be\label{potential}
V(\vec x)=\frac{1}{2}m\left[\omega^2_{\parallel}x_1^2+\omega^2_\perp(x_2^2+
x_3^2)\right]+A\cdot cos\left(2\pi\frac{x_1}{\sigma}\right)
\ee
The quadratic part is due to the interaction of the atoms with the magnetic 
field of the trap. According to laboratory implementations~\cite{Li97} we choose
$\omega_\parallel=2\pi\times 130 s^{-1}$ and $\omega_\perp=2\pi\times 150 s^{-1}$.
The additional term is generated by two opposite
laser beams in a standing wave configuration. A suitable choice of the
standing-wave parameters yields a double well-potential.
Taking into account the interatomic interaction, the atomic system is 
described by a macroscopic wave function $\psi$ that satisfies the 
GP equation
\be\label{grpi}
i\hbar\frac{\partial\psi}{\partial t}=
-\frac{\hbar^2}{2m}\nabla^2\psi+V\psi+g|\psi|^2\psi\equiv (H+g|\psi|^2)\psi,
\ee
with $H=-\hbar^2/(2m)\nabla^2+V$. 
Here, $m\simeq 7 a.u.$ is the mass of the lithium atom, and 
$g=(4\pi\hbar^2/m)a_s$, where $a_s$ is the s-wave scattering length for $^7Li$,
$a_s\simeq-1.45 nm$.
For a small number of atoms the GP nonlinearity 
can be neglected and Eq.~(\ref{grpi}) reduces to an ordinary Schr\"odinger 
equation. In such a case and for a sufficiently high barrier, the lowest 
energy level is described by a two-peak wave function symmetric with respect 
to inversion of the space 
axes, once the coordinate origin coincides with the trap center 
(Fig.~\ref{fig1}, dashed-dot line). 

Figure \ref{fig1} reports the spatial distribution of the ground 
state of 
Eq.~(\ref{grpi}) for different numbers of trapped atoms; we have used a numerical method
that consists in solving GP on a discrete space lattice and evaluating the lowest energy
state.
The barrier is specified by the two parameters 
$A_n\equiv A/\hbar=2650s^{-1}$ and $\sigma=5\mu m$ (see Eq.~(1)).
As shown in the figure, for $N=450$ the distribution is symmetric; 
instead for $N=500$ the nonlinear term is sufficient to destabilize the 
symmetric state, giving rise to two asymmetric stable states. 
For $N=600$ one well is almost empty.
As we increase the number of atoms, the nonlinearity plays a relevant role. 
By a self-consistent argument we realize that the symmetric wave function 
becomes unstable and we can have two new minimal energy states with             
distribution no longer symmetric for inversion (symmetry breaking). Indeed, 
let us assume a distribution as in Fig.~\ref{fig1} (dashed or solid line); then 
the effective potential for such a 
distribution, due to the sum of the external potential with $g|\psi|^2$ is an
asymmetric double well with the lower minimum corresponding to the higher
population peak.
For a sufficiently high nonlinear term the potential imbalance stabilizes the 
asymmetric distribution as in Fig.~\ref{fig1}. 

We confirm the numerical calculation by the following analytic model. Let $\psi_a$
be the equilibrium symmetrical wave function (either stable or unstable),
and $\psi_b$ be 
a suitable antisymmetrical wave function such that the weighted sum of the two 
wave functions lowers either one of the two peaks. In the two-dimensional space
of these wave functions, any other one can be expressed as 
\be\label{sottospazio}
\psi(\vec x)=a\psi_a(\vec x)+b\psi_b(\vec x).
\ee

Without loss of generality we can choose $\psi_a$ and $\psi_b$
as real functions and thus consider real values for $a$ and $b$. 
As we take $\int\psi_a^2d^3x=\int\psi_b^2d^3x=1$, it follows that
$a^2+b^2=\int\psi^2d^3x\equiv M(a,b)$, where $M(a,b)$, the total number
of atoms in the condensate, depends upon $a$ and $b$.
The energy is then given by
\be\label{hamilt}
{\mathcal H}=a^2H_{aa}+b^2H_{bb}+\frac{1}{2}(a^4I_{aa}+b^4I_{bb})+3a^2b^2I_{ab}
\ee
where 
$$
H_{aa}=\int\psi_a^*H\psi_ad^3x,\;\;\;\;
H_{bb}=\int\psi_b^*H\psi_bd^3x,
$$
$$
I_{aa}=g\int\psi_a^4d^3xx,\;\;\;\;
I_{bb}=g\int\psi_b^4d^3x,
$$
$$
I_{ab}=g\int\psi_a^2\psi_b^2d^3x.
$$
We look for the minimal of energy with the constraint of a fixed number $N$ of
condensed atoms.
These constrained minima satisfy the relations 
\be
\frac{\partial({\cal H}-\mu M)}{\partial a}=
\frac{\partial({\cal H}-\mu M)}{\partial b}=0,
\ee
where $\mu$ is a Lagrange multiplier. We thus solve for $a$ and $b$, with the
condition $N=a^2+b^2$.
The solutions $a=0$ and $b=0$ correspond, respectively, to the antisymmetric function
and the symmetric one. 
The other solutions, for both $a$ and $b$ nonzero, yield $a^2$, $b^2$ values as
functions of $\mu$. Using the constraint of fixed $N$, we eliminate $\mu$ and
find
\bey\label{2dimeq1}
a^2=\frac{H_{aa}-H_{bb}+N(3I_{ab}-I_{bb})}{6I_{ab}-I_{aa}-I_{bb}} \\
\label{2dimeq2}
b^2=\frac{H_{bb}-H_{aa}+N(3I_{ab}-I_{aa})}{6I_{ab}-I_{aa}-I_{bb}}
\eey
Since $g$ is negative,
the denominator is always negative. Indeed, $\psi_a^2$ and $\psi_b^2$
are almost equal at each point $x$.
For low $N$, the dominant terms in the 
numerators have opposite sign, thus one of the two squares has to  
be negative, which means that there is no solution with asymmetric wave 
function. 
On the other hand, for sufficiently large $N$, the second term in the numerator 
of the two equations can compensate for the positive one, since the quantities
$3I_{ab}-I_{aa}$, $3I_{ab}-I_{bb}$ are always negative.
Thus, we have proved that two asymmetrical steady states exist beyond a threshold value
of $N$.

In order to prove that the two states are stable, it is sufficient to show that the 
symmetrical state becomes unstable above threshold, that is, 
\be
\left.\frac{d^2\cal H}{d b^2}\right|_{b=0}\le0
\ee
where we have $a^2=N-b^2$. It is easily found that
\be\label{curvatura}
\left.\frac{d^2\cal H}{d b^2}\right|_{b=0}=2(H_{bb}-H_{aa})+2N(3I_{ab}-I_{aa})
\ee

Going back to Eq.~(\ref{2dimeq2}), the two stationary asymmetrical states occur when the
numerator changes sign. In fact, the numerator is the right side of Eq.~(\ref{curvatura}),
hence, at the critical point the symmetrical state becomes unstable.

One can easily evaluate the 
threshold value $N_i$ of $N$ for which this symmetry breaking occurs. 

\section{Macroscopic Quantum Coherence}

In such a bistable situation the energy displays two minima of equal value in 
the infinite-dimensional phase space. 
From a classical point of view, as the system is in its lowest energy state, 
the condensate is localized in either one of the two minima, where it will 
remain in the absence of thermal noise once we keep the atom number constant. 
Since, however, the condensate is a mesoscopic system, quantum fluctuations play 
a relevant role. This can be shown by second quantization of the field, 
replacing 
the {\cal c}-number macroscopic wave function by field operators. 

Quantum fluctuations 
allow the passage from one to the other stable state without thermal 
activation, by pure quantum tunneling. Furthermore, due to the coherent nature
of the process, we expect coherent oscillations between the two wells, that is,
MQC. We now evaluate the tunneling 
rate as a function of the system parameters showing the feasibility of MQC. 

The most natural way of evaluating the tunneling rate consists in finding the
two lowest eigenvalues of the Hamiltonian and taking their difference. 
Indeed, the sum and difference of the corresponding states are respectively the
alive and dead states of SC, and the transition time is 
half the period corresponding to the frequency difference.
\begin{figure}
\epsfig{figure=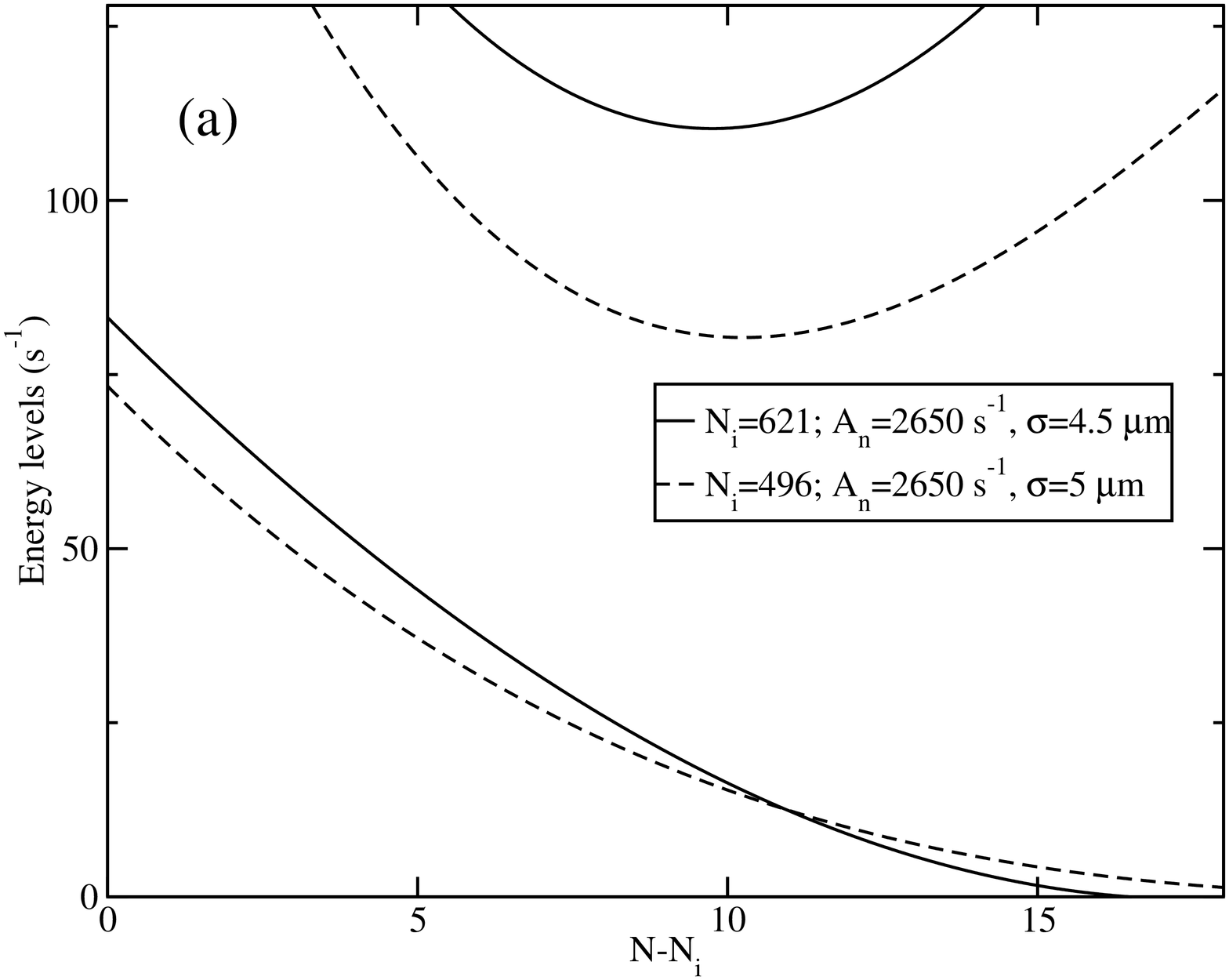,width=6cm,angle=-90}
\epsfig{figure=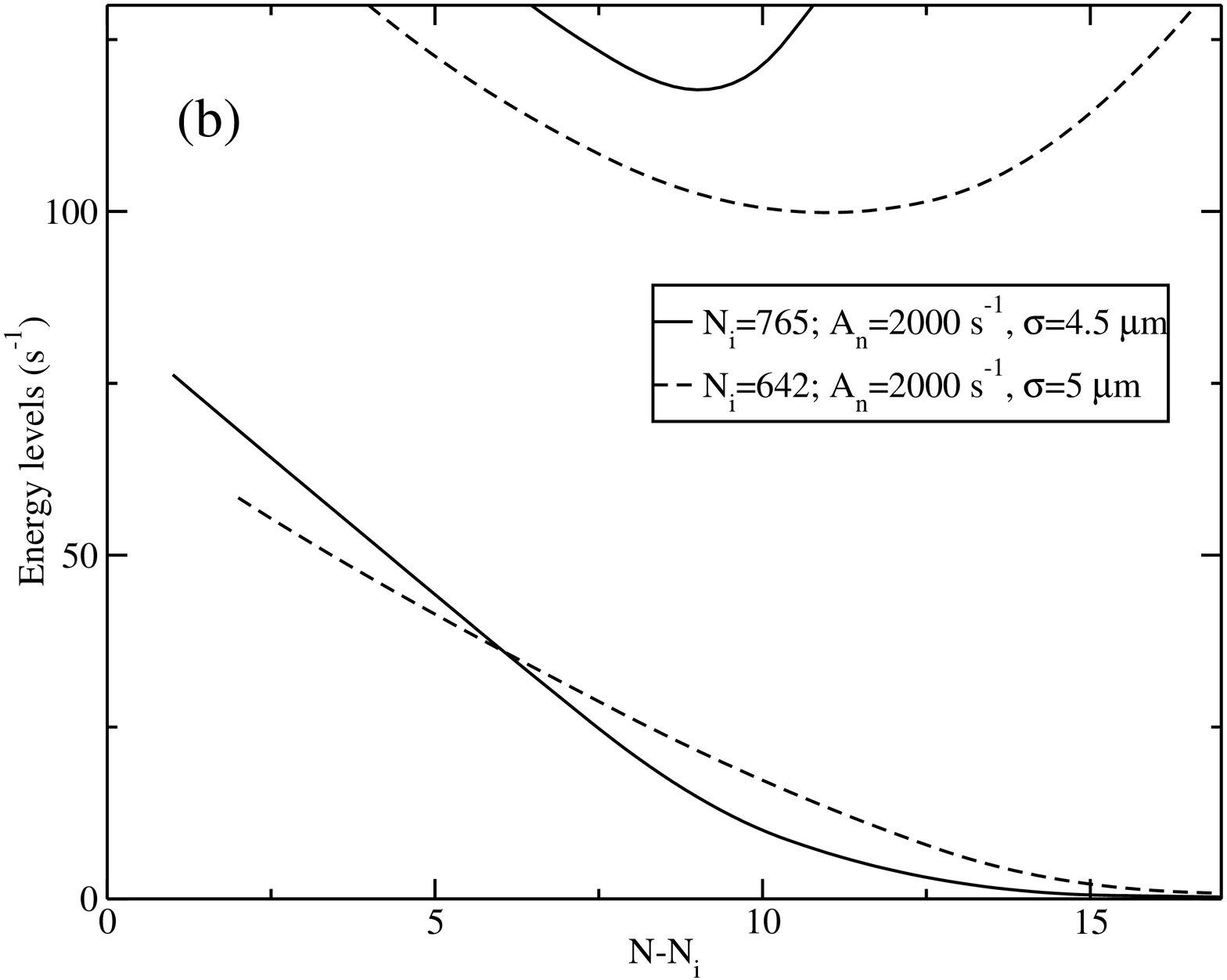,width=6cm,angle=-90}
\caption{First two excited energy levels versus the number $N-N_i$ of atoms above the
threshold $N_i$. Within each figure we keep the laser amplitude fixed and just 
vary the pitch of the potential lattice;
(a) corresponds to $A_n\equiv A/\hbar=2650 s^{-1}$ and
(b) corresponds to $A_n=2000 s^{-1}$.}
\label{fig2}
\end{figure}   

The problem is simplified by reducing it to two degrees of freedom by the 
expansion of Eq.(\ref{sottospazio}).

We select the basis functions $\psi_a$ and $\psi_b$ as follows. Calling 
$\psi_0(\vec x)$ the minimal-energy wave function of the GP problem (see e.g.
Fig.~\ref{fig1}), we take for $\psi_a$ and $\psi_b$, respectively, the symmetric and 
antisymmetric sums $\psi_0(\vec x)\pm\psi_0(-\vec x)$. Expansion (\ref{sottospazio})
with these $\psi_a$ and $\psi_b$ includes the original functions $\psi_0(\vec x)$
and $\psi_0(-\vec x)$ for suitable values of $a$ and $b$. Furthermore, it simplifies
the form of the Hamiltonian, as we see right now~\cite{nota}.

In second quantization, $a$ and $b$ in Eq.~(3) become the operators $\hat a$ and
$\hat b$, 
obeying Bose commutation rules with their conjugates $\hat a^\dag$, $\hat b^\dag$.
Exploiting the operator version of Eq.~(3) and its adjoint, the Hamiltonian becomes
\be\label{Hamiquant}
{\cal H}=\hat a^\dagger\hat aH_{aa}+\hat b^\dagger\hat bH_{bb}+\frac{1}{2}(
\hat a^{\dagger 2}\hat a^2I_{aa}+\hat b^{\dagger 2}\hat b^2I_{bb})+
\ee
$$
\left(\frac{1}{2}\hat a^{\dagger 2}\hat b^2+\frac{1}{2}\hat b^{\dagger 2}\hat a^2+2
\hat a^\dagger\hat b^\dagger\hat a\hat b\right) I_{ab}
$$

where the coefficients $H_{aa}$, $H_{bb}$, $I_{aa}$, $I_{bb}$ and $I_{ab}$ are the same 
as in Eq.~(4).

We consider the basis of eigenvectors of the number operators
\be
|0,N>, |1,N-1>,... |N,0>
\ee
where $\hat a^\dagger\hat a|k,m>=k|k,m>$ and $\hat b^\dagger\hat b|k,m>=m|k,m>$.
Let us call
\be
H_{l,k}=<l,N-l|{\cal H}|k,N-k>
\ee
the generic matrix element of the Hamiltonian on the above basis.
We evaluate the eigenvalues of this matrix. 
We have considered two different wavelengths and two amplitude values of the applied
field.

In Fig.~\ref{fig2} we report on the first two excited energy levels versus the number
of condensed atoms beyond the threshold value. In such figure we keep constant the barrier
height and just vary the barrier width. Notice that for increasing $\sigma$ the maximum
tunneling frequency reduces, but the slope at which it reduces for increasing $N-N_i$ is
less steep.

\begin{figure}
\epsfig{figure=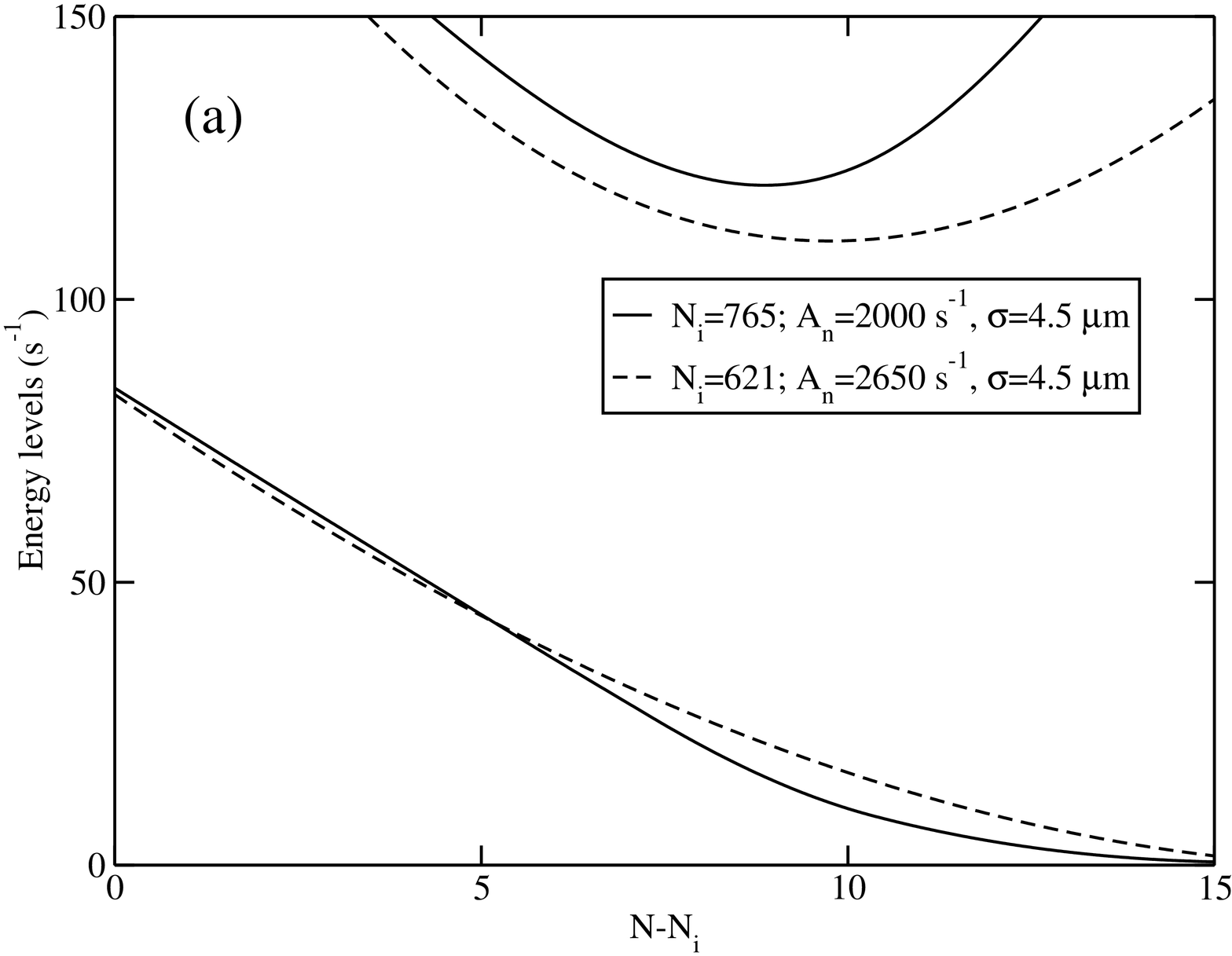,width=6cm,angle=-90}
\epsfig{figure=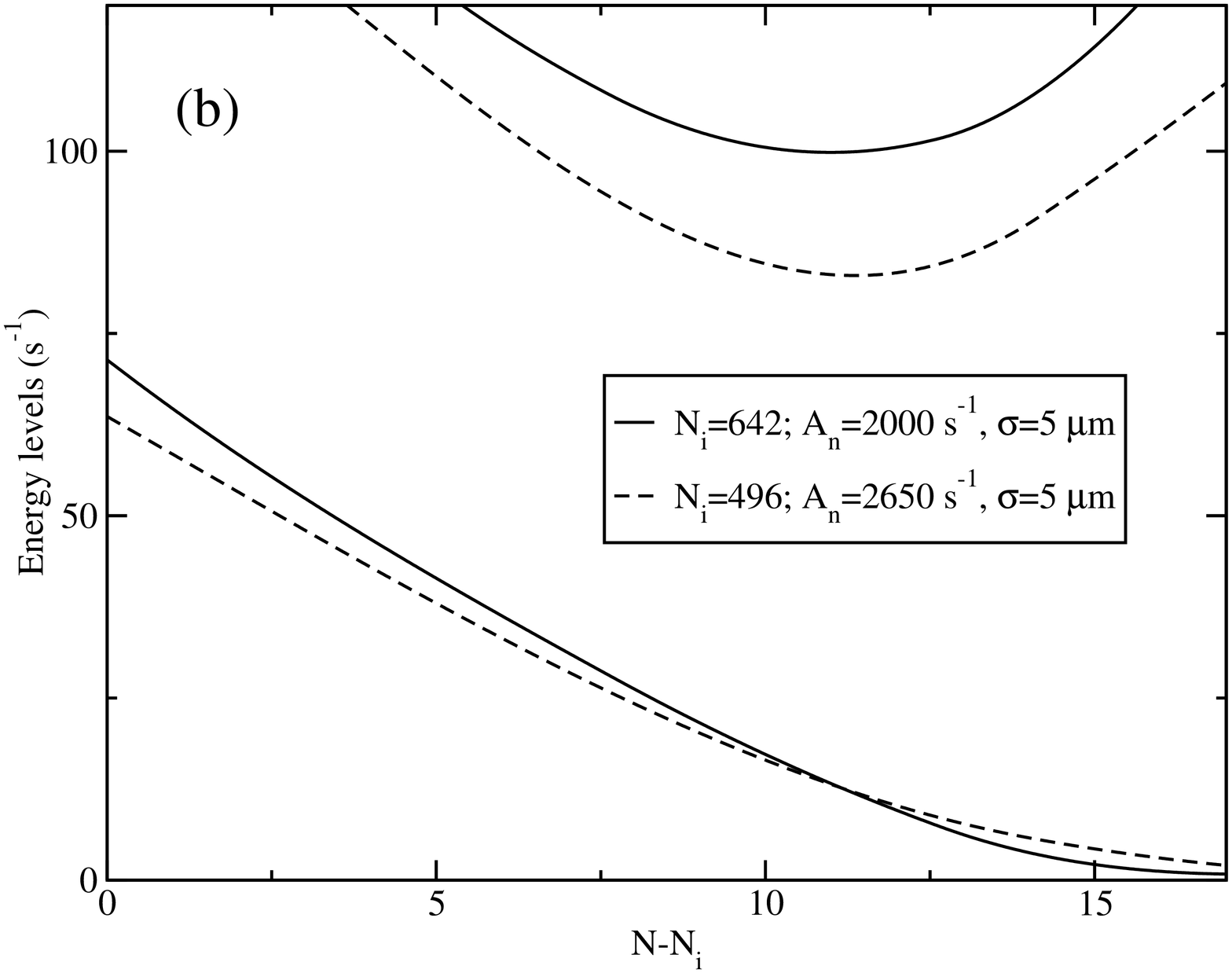,width=6cm,angle=-90}
\caption{Same as with Fig.~\ref{fig2}, but with fixed $\sigma$, respectively (a) 
$\sigma=4.5\mu m$ and (b) $\sigma=5\mu m$.}
\label{fig3}
\end{figure}

In Fig.~\ref{fig3} we keep fixed the barrier width and change its height.
Here too the maximum frequency decreases for increased heights, but again as in
Fig.~\ref{fig2}, the slope decreases for increasing $N-N_i$.

\begin{figure}
\epsfig{figure=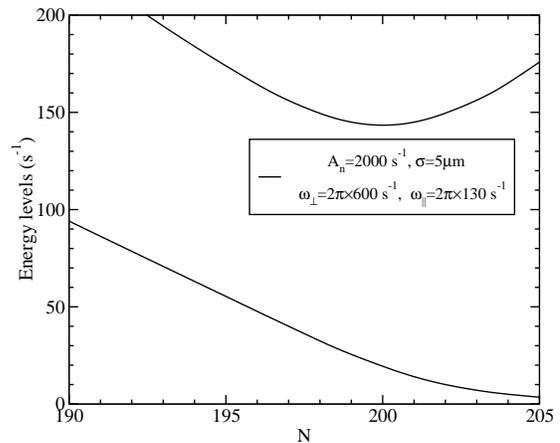,width=6cm,angle=-90}
\caption{First two excited energy levels versus the number of atoms $N$.
The system parameters are $\omega_\bot=2\pi\times600 s^{-1}$, 
$\omega_\parallel=2\pi\times130 s^{-1}$,
$A_n\equiv A/\hbar=2000 s^{-1}$ and $\sigma=5 \mu m$.}
\label{fig4}
\end{figure}

In Fig.~\ref{fig4} we reduce the threshold for BEC breaking by increasing $\omega_\bot$.
Precisely, we report on the first two excited states for 
$\omega_\bot=2\pi\times600 s^{-1}$,
$A_n=2000 s^{-1}$ and $\sigma=5\mu m$. The threshold value is $N_i\simeq190$. 

\begin{figure}
\epsfig{figure=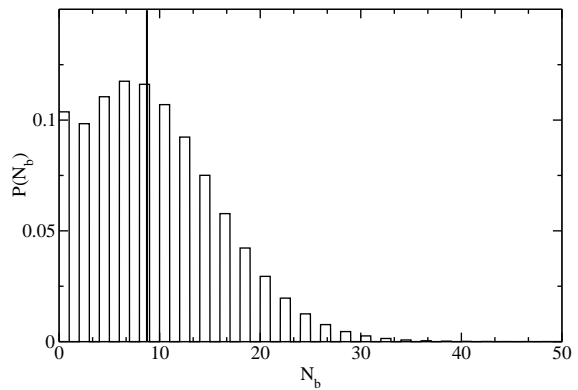,width=6cm,angle=-90}
\caption{Probability distribution of the population in the ground state as a function
of the number of atoms in mode 'b'. $A=2000 s^{-1}$, $\sigma=5\mu m$ and $N=655$.}
\label{fig5}
\end{figure}

In the number representation there is no explicit evidence of a SC
as a two-peak distribution (see Fig.~\ref{fig5}). We look for a suitable 
observable, whose probability distribution provides such an evidence. We take 
the first component of the barycenter coordinate $x_c=(1/N)\int x_1|\psi|^2d^3x$ as the 
appropriate variable since the corresponding classical states [minima of the 
Hamiltonian~(\ref{hamilt})] have separated barycenters.
It is associated with the operator

\be
\hat x_c=\frac{1}{N}\int x_1\psi_a(\vec x)\psi_b(\vec x)d^3x
(\hat a^\dagger\hat b+\hat a\hat b^\dagger)
\ee
Besides a c-number factor, the observable is thus $\hat M=\hat a^\dagger\hat b+\hat 
a\hat b^\dagger$. The associated probability density is $P(m)=|<m|\phi_0>|^2$, where 
$|\phi_0>$ is the ground state of Hamiltonian (\ref{Hamiquant}) and $|m>$ is the 
eigenstate of $\hat M$ with eigenvalue $m$.
Notice that once the number of atoms is fixed the eigenvectors are not degenerate.

Since we know the components of $|\phi_0>$ on the number basis we must express $|m>$ with 
respect to that basis, that is, 
\be
|m>=\sum_k c_k^m|k,N-k>.
\ee
Applying the operator $\hat M$ to the above ket we have
\be
m|m>=\sum_k c_k^m(\hat a^\dagger\hat b+\hat a\hat b^\dagger)|k,N-k>.
\ee
Projecting on $<l,N-l|$ the above ket it follows that 
\be\label{eigenv}
\sum_k M_{l,k}c_k^m=m c_l^m,
\ee
where 
\be
M_{l,k}=<l,N-l|\hat M|k,N-k>.
\ee
\begin{figure}
\epsfig{figure=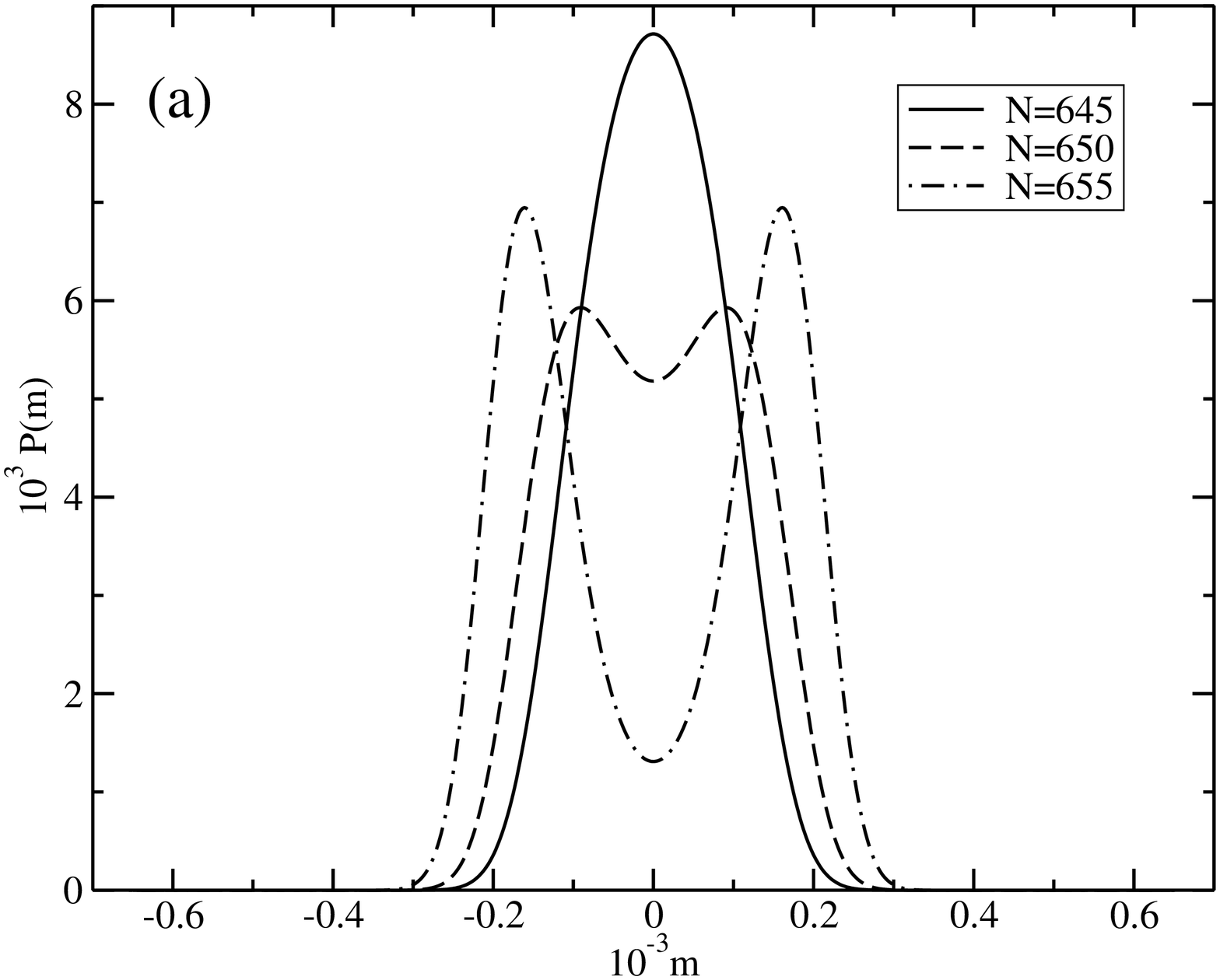,width=7cm,angle=-90}
\epsfig{figure=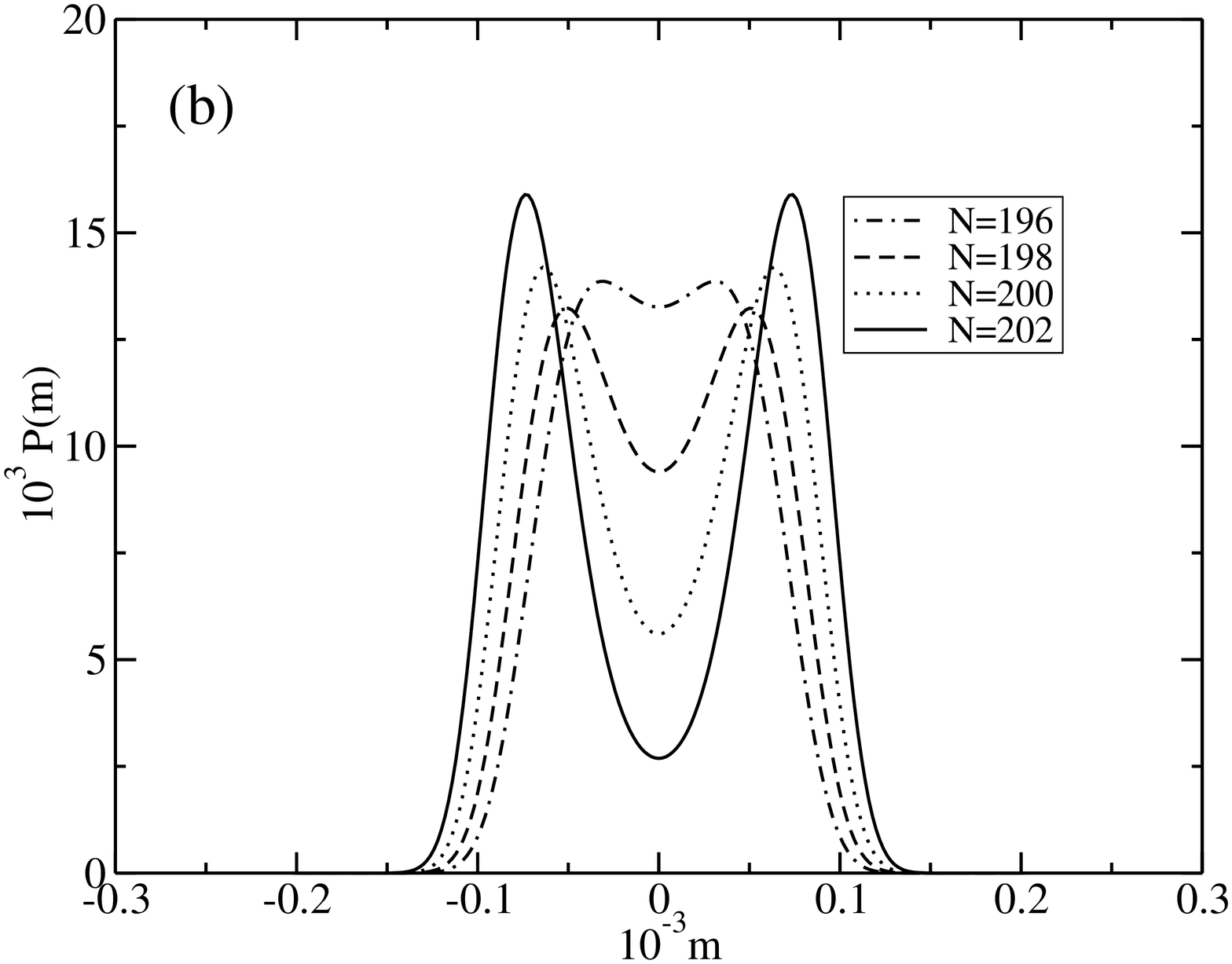,width=7cm,angle=-90}
\caption{Probability distribution of the "barycenter" indicator for the lithium BEC,
for different $N$ values and for a standing wave with 
$A_n=2000 s^{-1}$ and $\sigma=5\mu m$; (a) $\omega_\bot=2\pi\times150 s^{-1}$ and
(b) $\omega_\bot=2\pi\times600 s^{-1}$.}
\label{fig6}
\end{figure}

Since $M_{l,k}$ are known, solving the eigenvector equation (\ref{eigenv}) we can evaluate 
the coefficients $c_k^m$ and hence $P(m)$.

The two-peak distribution $P(m)$ is plotted in Fig.~\ref{fig6}(a) for different $N$ values and 
for $A/\hbar=2000s^{-1}$, $\sigma=5\mu m$. The same distributions, but for 
$\omega_\perp=2\pi\times600 s^{-1}$, are given in Fig.~\ref{fig6}(b).

\begin{figure}
\epsfig{figure=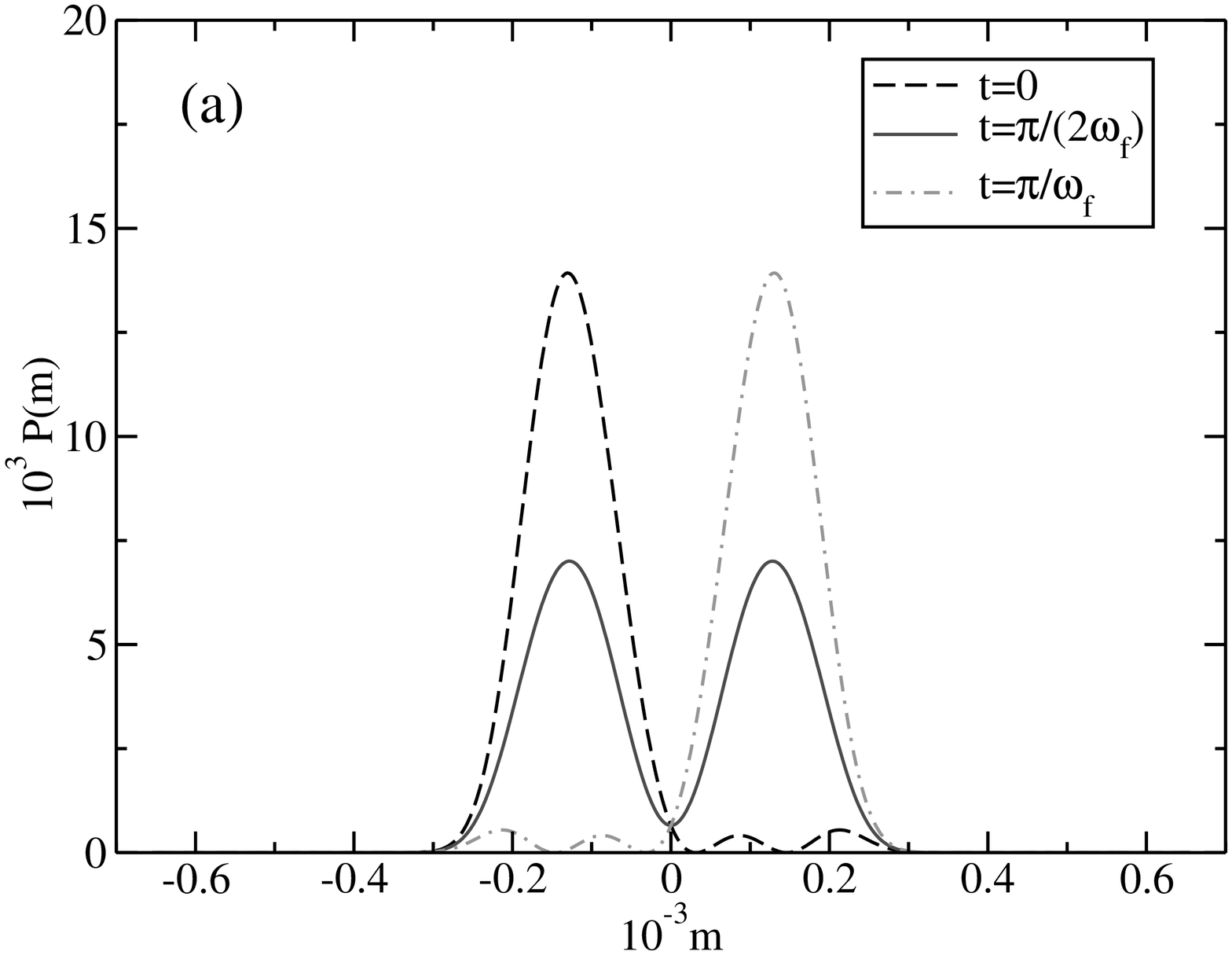,width=6cm,angle=-90}
\epsfig{figure=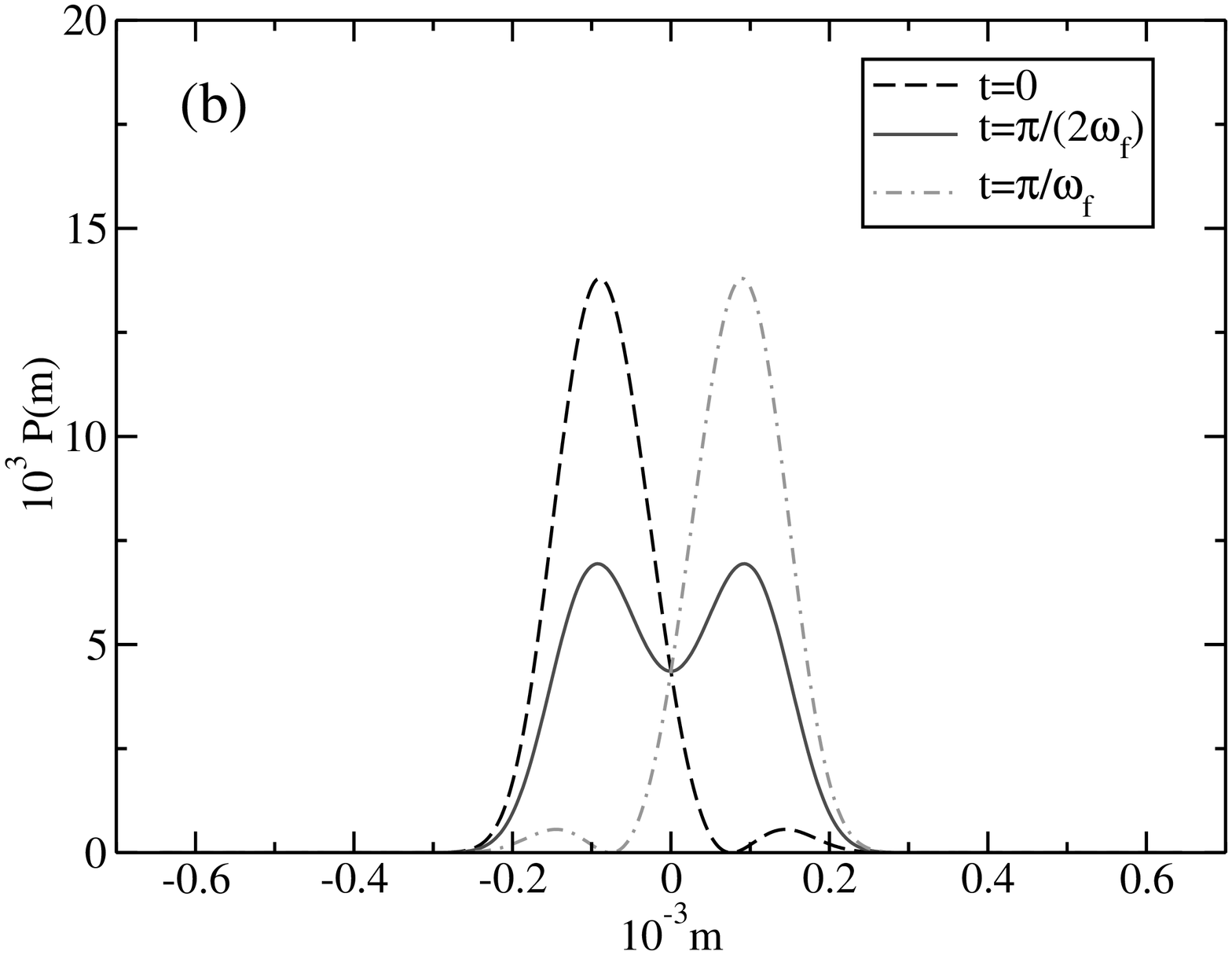,width=6cm,angle=-90}
\caption{Time evolution of the probability distribution. $\omega_f$ is the tunneling
frequency. The parameters of the standy wave are as in 
Fig.~\ref{fig6}(a); (a) number of atoms $N=655$ and (b) $N=645$, for which the ground state
does not display two separate peaks [see solid line of Fig.~\ref{fig6}(a)].}
\label{fig7}
\end{figure}

Experimentally, evidence of SC against a trivial statistical mixture is obtainable by 
setting the system at the above parameter values and observing the coherent oscillation 
between the two energy minima by a measurement not destroying the coherence 
(e.g., a phase contrast technique). 
The condensate can be prepared in the ground state, because
the system condensates naturally in this state. Then one measures
the condensate barycenter by the above noninvasive technique
at the initial time. This
measurement collapses the distribution $P(m)$ onto one peak. If
the energy transfer of the measurement is not too large,
we expect that only the first excited state is populated and
hence we have the superposition of $\phi_0(m)$ and $\phi_1(m)$
[see Fig.~\ref{fig7}(a)]. 
At the following times the distribution oscillates between
the two peaks.
The symmetry of Hamiltonian (\ref{Hamiquant}) with respect to the transformation 
$\hat a\rightarrow -\hat a$
shows that $\phi_0(m)$ and $\phi_1(m)$ are symmetric or antisymmetric; by a suitable
choice of the phase, both $\phi_0$ and $\phi_1$ can be real functions. We consider a state
preparation such that at $t=0$ the state $\phi_0(m)+\phi_1(m)$ has only one probability
peak [see Fig.~\ref{fig7}(a), dashed line].
After a quarter period corresponding to the frequency separation $\omega_f$ between ground 
and first excited states, the superposition will be $\phi_0(m)-i\phi_1(m)$ and the 
corresponding probability is the two-peak aspect (solid line),
\be\label{pippo}
Q(m,t=\pi/\omega_f)=\phi_0(m)^2+\phi_1(m)^2.
\ee

At time $t=\pi/\omega_f$ the only peak is that missing at time $t=0$, thus there is a
coherent oscillation between the two states.
Detecting such an oscillation would provide evidence of an SC at an intermediate time
when both peaks are present.

The intermediate distribution, given by Eq.~(\ref{pippo}), may display 
two peaks even when the ground state has only one [see Fig.~\ref{fig7}(b)]. This occurs
because the distribution $\phi_1(m)^2$ has always two peaks.

Notice that if we had $N$ loosely coupled or independent atoms
($Ng$ sufficiently low or even zero) the superposition of ground
and first excited state would have a single peak, oscillating
as a coherent state inside a harmonic potential. This would
by no means be a SC. On the contrary, we have shown that for
$Ng$ sufficiently high, we have a two peak distribution with
the two partial barycenters at nearly constant positions.
During the evolution, the two peak amplitudes oscillate, that
is, the probability to find the system in either state
oscillates.

In conclusion, we have found that the oscillation frequency between
the two states of the SC is $\omega_f\sim 50s^{-1}$ [see Fig.~\ref{fig2}(a) for $N-N_i\sim5$]. 
In order to neglect thermal activation, the system should be cooled at
a temperature of around $1 nK$. To cool at $1 nK$ is within the reach of present 
technologies, even though such
a low temperature has not yet been reported. On the other hand, the collective degrees of 
freedom for which there is quantum coherence may be weakly coupled with the other modes
of the condensate, which act as a thermal bath. Hence, in such a case it might not be
necessary to cool at $1 nK$ the whole condensate, but just the involved degrees of
freedom. 
 
\section{Losses and observability of MQC}

Besides the finite temperatures effects, we must also account for atom losses from the
condensate.

The entanglement between condensate and lost atoms will imply a decoherence of the
superposition. 
Near the threshold for symmetry breaking, the two macroscopic wave functions corresponding
to the energy minima of the classical system are almost coincident (see Fig.~\ref{fig1},
for $N=500$), thus the loss of one or a few atoms does not allow us yet to discriminate
between the two alternatives, and thus decoherence effects are almost negligible.

We perform a simple calculation of the decoherence effect due to atom loss for the 
condensate. 
In the Hartree-Fock approximation all the atoms are in the same state, given by the 
GP wave function. Hence, if $\psi_0(\vec x)$ is one of the two asymmetrical states with
minimal energy, the wavefunction for the whole system of $N$ atoms is given by
\be
\psi_l(\vec x_1,\vec x_2,...,\vec x_N)=\prod_{i=1}^{N}\psi_0(\vec x_i).
\ee
But the GP yields another minimum $\psi_0(-\vec x)$, for which the overall wave function
is given by
\be
\psi_r(\vec x_1,\vec x_2,...,\vec x_N)=\prod_{i=1}^{N}\psi_0(-\vec x_i).
\ee
The coherent superposition of the two alternatives is given by
\be
\psi_s=K(\psi_l+\psi_r)
\ee
where $K$ is a normalization factor. If $\psi_{l,r}$ are orthogonal then 
$K=\frac{1}{\sqrt{2}}$.  

\begin{figure}
\epsfig{figure=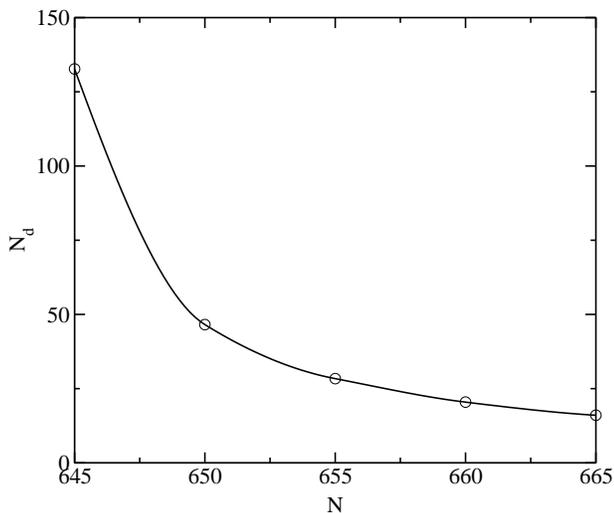,width=7cm,angle=-90}
\caption{Number $N_d$ of lost atoms for which the coherence decays by $1/e$ versus
the total number of atoms, for
$A_n\equiv A/\hbar=2000 s^{-1}$ and $\sigma=5 \mu m$.}
\label{fig8}
\end{figure}

The density operator is
\be
\hat\rho=\frac{1}{2}\left(|\psi_l\rangle+|\psi_r\rangle)(\langle\psi_l|+\langle\psi_r|
\right)
\ee
The coherence between the two alternatives is given by 
\be
C=2 Tr[|\psi_l\rangle\langle\psi_r|\hat\rho]=1.
\ee
If we assume that the atoms escaping the condensate transfer a negligible energy to the
trapped atoms, then the residual coherence after the loss of $M$ atoms is
\be
\tilde C=2 Tr[|\tilde\psi_l\rangle\langle\tilde\psi_r|\hat\rho],
\ee
where the vectors $|\tilde\psi_{l,r}>$ refer to the $N-M$ atoms still in the condensate.
The functions $\psi_{l,r}$ can be written as
\be
\psi_{l,r}=\tilde\psi_{l,r}\prod_{i=1}^{M}\psi_0(\pm\vec x_i),
\ee
where $i=1,2,...,M$ are the lost atoms. We rewrite $\hat\rho$ by using this expansion,
and find, provided that $\langle\tilde\psi_l|\tilde\psi_r\rangle=0$,  
\be
\tilde C=\left(\int d^3x\psi_0(\vec x)\psi_0(-\vec x)\right)^M\equiv (I_o)^M.
\ee
Let us call $\epsilon=1-I_o$. If $\epsilon\ll 1$ then the coherence is given by
\be
\tilde C\simeq e^{-\epsilon M}.
\ee
The quantity $N_d=1/\epsilon$ provides the number of atoms that must be lost in order
to reduce the coherence by $1/e$. Figure~\ref{fig8} shows how $N_d$ scales with the
total number of atoms $N$ for $A=2000 s^{-1}$ and $\sigma=5\mu m$.

Notice that near the threshold value ($N=643$) $N_d$ is rather large, and hence,
coherence is more robust with respect to atom losses. Far threshold $N_d$ has a
weak dependence on $N$.

We now evaluate the loss rate of the condensate. The relevant processes are two-body
inelastic and three-body collisional decays. 
The total loss rate of the two decays are~\cite{Dodd}
\be
R(N)=\alpha N^2\int d^3r|\psi_0(r)|^4+L N^3\int d^3r|\psi_0(\vec x)|^6
\ee
where $\alpha$ is the two-body dipolar loss rate coefficient and $L$ is the three-body 
recombination loss rate coefficient. We use the following values: $\alpha=1.2\times10^{-14}
cm^3s^{-1}$~\cite{Dodd}, $L=2.6\times10^{-28}cm^6s^{-1}$ \cite{Moe}.

\begin{figure}
\epsfig{figure=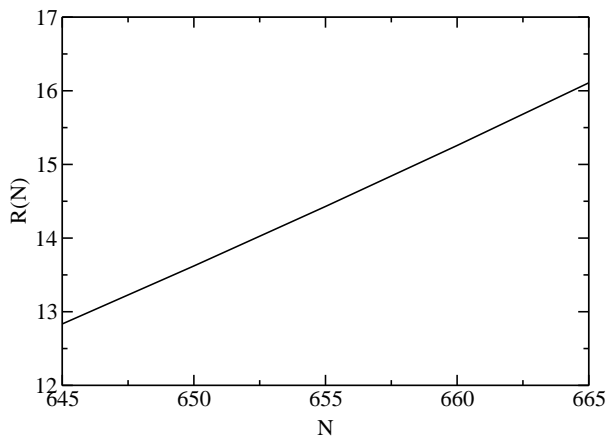,width=7cm,angle=-90}
\caption{R(N) for
$A_n\equiv A/\hbar=2000 s^{-1}$ and $\sigma=5 \mu m$.}
\label{fig9}
\end{figure}

We report in Fig.~\ref{fig9} $R(N)$ for $A=2000 s^{-1}$ and $\sigma=5 \mu m$. From 
Figs.~\ref{fig2}(b),\ref{fig8}, and \ref{fig9} we can infer that near threshold atom
losses have a small effect on coherence. 

For instance, for $N=650$ the tunneling rate
is about $25 s^{-1}$ [see Figs.~\ref{fig2}(b) and \ref{fig3}(b)], thus the number of 
atoms lost during an oscillation period 
($\sim0.25 s^{-1}$) is around $3.5$ (see Fig.~\ref{fig9}); but Fig.~\ref{fig8} shows 
that such a
loss is negligible for coherence. For $N=655$ the tunneling rate is about $10 s^{-1}$,
thus the number of atoms lost during an oscillation period is about $9$, still below 
$N_d\simeq30$.

Besides decoherence, the atom loss yields a shift in the tunneling frequencies.
As shown in Figs.~\ref{fig2}-\ref{fig3}, the frequencies vary a lot as we reduce the
number of condensed atoms. For $N=650$ the lost atoms are around $3.5$ during an 
oscillation period, thus the oscillation frequency changes by $\sim30\%$.

\begin{figure}
\epsfig{figure=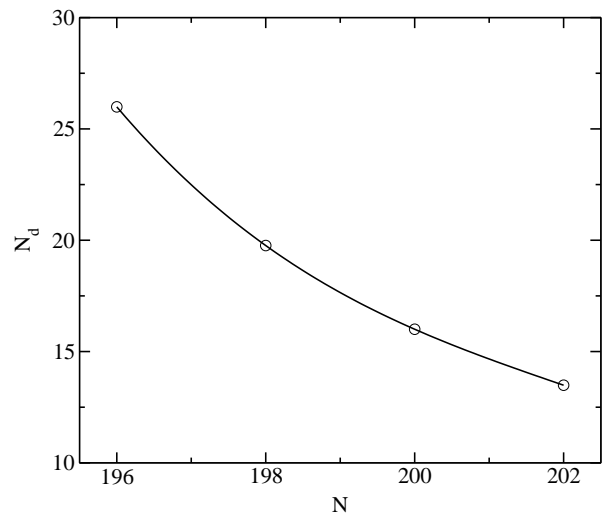,width=7cm,angle=-90}
\caption{Number $N_d$ of lost atoms for which the coherence decays by $1/e$ versus
the total number of atoms, for the parameters of Fig.~\ref{fig4}.}
\label{fig10}
\end{figure}

\begin{figure}
\epsfig{figure=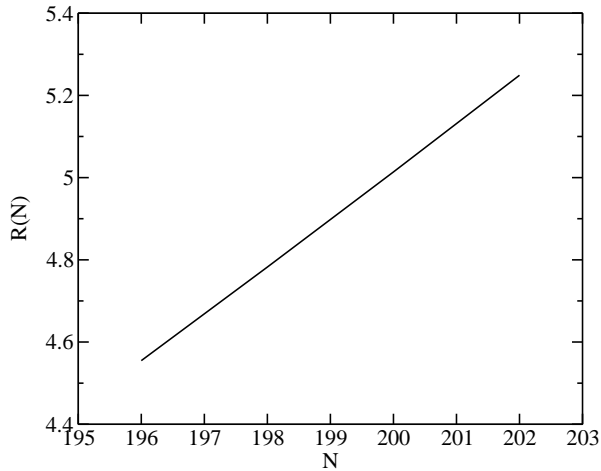,width=7cm,angle=-90}
\caption{R(N) for the parameters of Fig.~\ref{fig4}.}
\label{fig11}
\end{figure}

In order to reduce the effect of losses we lower the threshold value by increasing
the frequency $\omega_\bot$. In Figs.~\ref{fig10} and \ref{fig11} we report on 
$N_d$ and $R(N)$ for $\omega_\bot=2\pi\times600 s^{-1}$; the longitudinal frequency and 
the standing
wave have been kept unchanged. The tunneling frequencies and probabilities $P(m)$ are
given in Figs.~\ref{fig4}(b) and~\ref{fig6}(b), respectively.
We notice that here the loss rate is much smaller, and hence the frequency shift is smaller, 
whereas the tunneling frequencies
are slightly larger. A further reduction of the threshold value should improve the
situation.

\section{Conclusion}

We have solved numerically the Gross-Pitaevskii equation for a $^7Li$ condensate in
a double-well potential and we have shown that the spatial density undergoes a symmetry
breaking for a suitable number of condensed atoms. The classical asymmetrical wave function
is used to select two modes. With their quantization we have evaluated the tunneling rate for
some parameters of the system and we have found that it is within the reach of present
laboratory technologies. Thermal effects could be lower than expected because of a
decoupling between the collective variables and the thermal bath.
To prove the presence of two alternatives we have introduced an appropriate observable.
We have then evaluated the effects of loss and we have showed 
that the decoherence is negligible. In a forthcoming work we discuss the conditions 
for bistability in the case of mutual repulsion (positive scattering length).


\begin{thebibliography}{20}
\bibitem{RbCond} In the case of positive scattering length as for $^{87}Rb$~\cite{RbRef} 
or $^{23}Na$~\cite{Ketterle},
the number of atoms in a condensate may be as large as $10^6$ or $10^7$, respectively.
\bibitem{nota} A generic two mode model is also used by C.~Menotti et al., Phys. Rev. A
{\bf 63}, 023601 (2001), however, here we have selected the suitable modes by 
reference to the associated GP equation.
\end{thebibliography}
\end{document}